\documentclass[sigconf,nonacm]{acmart}

\usepackage{graphicx}
\usepackage{subfiles}

\usepackage{todonotes}
\usepackage{booktabs}
\PassOptionsToPackage{hyphens}{url}\usepackage{hyperref}

\usepackage{dsfont}
\usepackage{amsmath}

\usepackage{caption}
\usepackage{subcaption}
\usepackage{tcolorbox}

\usepackage{xcolor}

\AtBeginDocument{%
  \providecommand\BibTeX{{%
    \normalfont B\kern-0.5em{\scshape i\kern-0.25em b}\kern-0.8em\TeX}}}

\setcopyright{acmcopyright}
\copyrightyear{2018}
\acmYear{2018}
\acmDOI{10.1145/1122445.1122456}

\acmConference[Woodstock '18]{Woodstock '18: ACM Symposium on Neural
  Gaze Detection}{June 03--05, 2018}{Woodstock, NY}
\acmBooktitle{Woodstock '18: ACM Symposium on Neural Gaze Detection,
  June 03--05, 2018, Woodstock, NY}
\acmPrice{15.00}
\acmISBN{978-1-4503-XXXX-X/18/06}

\begin{abstract}

The HLTCOE LiveRAG submission utilized the GPT-researcher framework for researching the context of the question, filtering the returned results, and generating the final answer. The retrieval system was a ColBERT bi-encoder architecture, which represents a passage with many dense tokens. Retrieval used a local, compressed index of the FineWeb10-BT collection created with PLAID-X, using a model fine-tuned for multilingual retrieval.
Query generation from context was done with Qwen2.5-7B-Instruct, while filtering was accomplished with m2-bert-80M-8k-retrieval. Up to nine passages were used as context to generate an answer using Falcon3-10B. This system placed 5th in the LiveRAG automatic evaluation for correctness with a score of 1.07.
\end{abstract}

\begin{document}

\title{HLTCOE at LiveRAG: GPT-Researcher using ColBERT retrieval}

\author{Kevin Duh}

\affiliation{
  \institution{HLTCOE at Johns Hopkins University}
  \city{Baltimore}
  \state{MD}
  \country{USA}
}
\email{kduh1@jhu.edu}
\author{Eugene Yang}

\affiliation{
  \institution{HLTCOE at Johns Hopkins University}
  \city{Baltimore}
  \state{MD}
  \country{USA}
}
\email{eugene.yang@jhu.edu}
\author{Orion Weller}

\affiliation{
  \institution{Johns Hopkins University}
  \city{Baltimore}
  \state{MD}
  \country{USA}
}
\email{oweller2@jhu.edu}
\author{Andrew Yates}

\affiliation{
  \institution{HLTCOE at Johns Hopkins University}
  \city{Baltimore}
  \state{MD}
  \country{USA}
}
\email{andrew.yates@jhu.edu}
\author{Dawn Lawrie}

\affiliation{
  \institution{HLTCOE at Johns Hopkins University}
  \city{Baltimore}
  \state{MD}
  \country{USA}
}
\email{lawrie@jhu.edu}

\renewcommand{\shortauthors}{Kevin Duh, et al.}





\maketitle



\begin{figure}
  \centering
  \vspace{1em}
  \includegraphics[width=\linewidth]{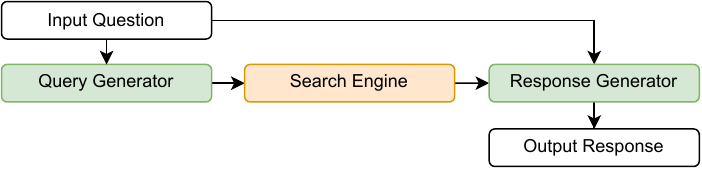}
  \caption{System design for the HLTCOE RAG system. }
  \label{fig:design}
\end{figure}

\section{Introduction}

We adopted GPT-Researcher as the main framework for our LiveRAG submission to use an abstractive summarization paradigm to support answer generation. This approach was one of the best performing systems in TREC NeuCLIR's report generation tasks~\cite{neuclir2024}.
We used Qwen2.5-7B-Instruct\footnote{https://huggingface.co/Qwen/Qwen2.5-72B-Instruct}~\cite{qwen2025qwen25technicalreport} with 8 bit quantization for query formulation.
We used a ColBERT-based search engine~\cite{ecir2022colbert-x} for retrieval with a PLAID-X model trained multilingually~\cite{ecir2024translate-distill}. The retrieval system returned top-ranked passages. Passages were then chunked into snippets. 
Passage snippets were filtered based on the cosine similarity between the original question and the snippet using m2-bert-80M-8k-retrieval\footnote{https://huggingface.co/togethercomputer/m2-bert-80M-8k-retrieval}~\cite{fu2023monarch}. The general architecture appears in Figure~\ref{fig:design}.
Finally, Falcon3-10B-Instruct\footnote{https://huggingface.co/tiiuae/Falcon3-10B-Instruct}~\cite{Falcon3} was used to generate the final response. The remainder of the paper outlines the detailed architecture of our system as well the prompts we used in the different steps in our pipeline. We describe the steps we took to validate the approach and optimize the performance to be able to perform at the required speed for the competition.

\section{System Design}

Our RAG system divided the process into two main stages. In the first stage research was conducted to identify relevant information and in the second stage the response was written. Conducting research included generating queries based on an initial search and then gathering all the responsive data together from the retrieved passages.
Writing the response used the gathered context to generate the final output of the system. In Section~\ref{sec:research} we describe the process for conducting research. Section~\ref{sec:write}
describes the process of generating the response given the research.

\subsection{Conduct Research}
\label{sec:research}

The research stage involves four major steps:
(1) issue an initial retrieval to set the context, (2) review the context to determine additional search queries, (3) issue additional search queries, and (4) divide passages into snippets and the snippets based on responsiveness to the original LiveRAG question.

To set the initial context for the research, the top three ranked passages are used based on the query entered by the user, in this case the content of the {\tt{question}} field for LiveRAG. Retrieval was done with an in-house index of the FineWeb-10BT collection. A ColBERT architecture search engine~\cite{ecir2022colbert-x} was used, which creates a dense vector per token for both the passage and query. We divided documents into non-overlapping passages of 450 tokens each for indexing. For efficiency, the index was compressed using the PLAID-X version\footnote{\url{https://pypi.org/project/PLAID-X/}} of ColBERT. Since it was possible for FineWeb-10BT to contain non-English documents, we chose to use an XLMR model\footnote{https://huggingface.co/FacebookAI/xlm-roberta-large}~\cite{conneau2019unsupervised} fine-tuned using translate distill~\cite{ecir2024translate-distill} for multilingual retrieval, which is trained with English queries and documents in several languages including English and Spanish. The training dataset is MS~MARCO~\cite{msmarco}. MS~MARCO documents were translated in-house with Sockeye Version~2~\cite{sockeye2} to create the training data.

The prompt in Figure~\ref{fig:research_prompt} was used to determine additional queries that were used for retrieval. We set fields referenced in the prompt as follows:
\begin{itemize}
\item {\tt{max\_iterations}} - 2
\item {\tt{task}} - the value of the {\tt{question}} field
\item {\tt{date}} - the date the software is run
\item {\tt{context}} - a concatenation of the top three ranked passages using the {\tt{question}} as the search query
\item {\tt{dynamic\_examples}} - based on {\tt{max\_iterations}} and in this case is "query 1", "query 2"
\end{itemize}
This prompt was issued to the Qwen2.5-7B-Instruct-Turbo model.  The top three passages were kept from each of the two queries generated as well as the original query. This created a list of nine passages. 
Our PLAID-X search engine was used to return passages for each the queries.

The final step of conducting research utilized the m2-bert-80M-8k-retrieval model to embed both the original {\tt{question}} and the passage in chunks of 1000 characters with 100 character overlap, referred to as snippets herein. 
The cosine similarity of the two vectors was computed. Any snippet whose similarity with the {\tt{question}} was at least 0.35 was maintained for answer generation.


\begin{figure}[tb]

\begin{tcolorbox}[width=\linewidth, left=0.5em, right=0.5em]
Write \{max\_iterations\} google search queries to search online that form an objective opinion from the following task: "\{task\}"

Assume the current date is \{date\} if required.
You are a seasoned research assistant tasked with generating search queries to find relevant information for the following task: "\{task\}".
Context: \{context\}
 
Use this context to inform and refine your search queries. The context provides real-time web information that can help you generate more specific and relevant queries. Consider any current events, recent developments, or specific details mentioned in the context that could enhance the search queries.

You must respond with a list of strings in the following format: {[$\{$dynamic\_example$\}$]}.
The response should contain ONLY the list.
\end{tcolorbox}
  \caption{Query Generation Prompt}
  \label{fig:research_prompt}
\end{figure}

\subsection{Generate Response}
\label{sec:write}
\begin{figure}[tb]

\begin{tcolorbox}[width=\linewidth, left=0.5em, right=0.5em]
Information: "\{context\}"
---
Using the above information, answer the following query or task: "\{question\}" in one or two sentences.
Use at most \{total\_words\} words.
\end{tcolorbox}
  \caption{Response Generation Prompt}
  \label{fig:write_prompt}
\end{figure}

\begin{figure*}
\begin{centering}
    \begin{subfigure}[t]{0.33\textwidth}
    \includegraphics[scale=0.45]{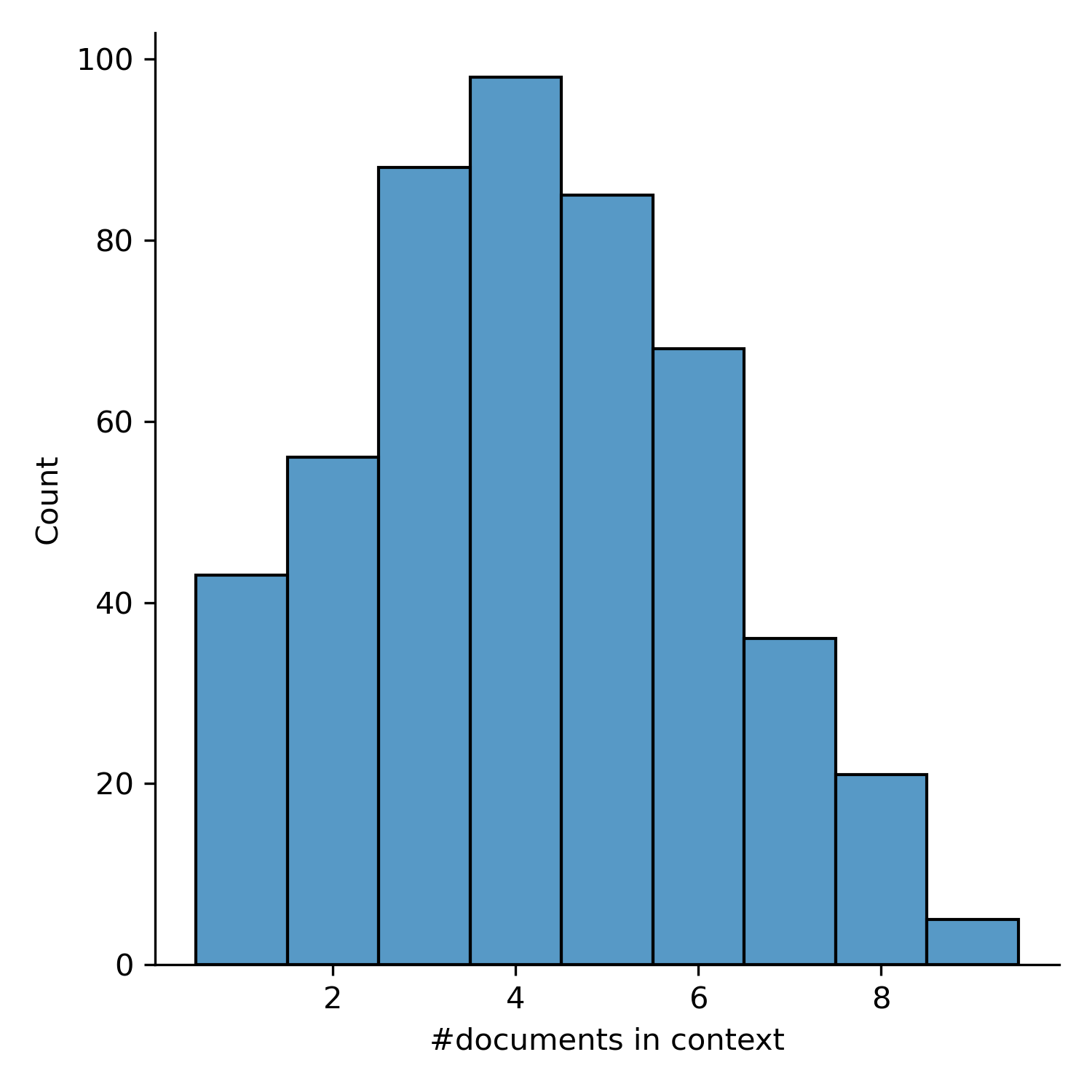}
    \end{subfigure}%
    ~ 
    \begin{subfigure}[t]{0.33\textwidth}
    \includegraphics[scale=0.45]{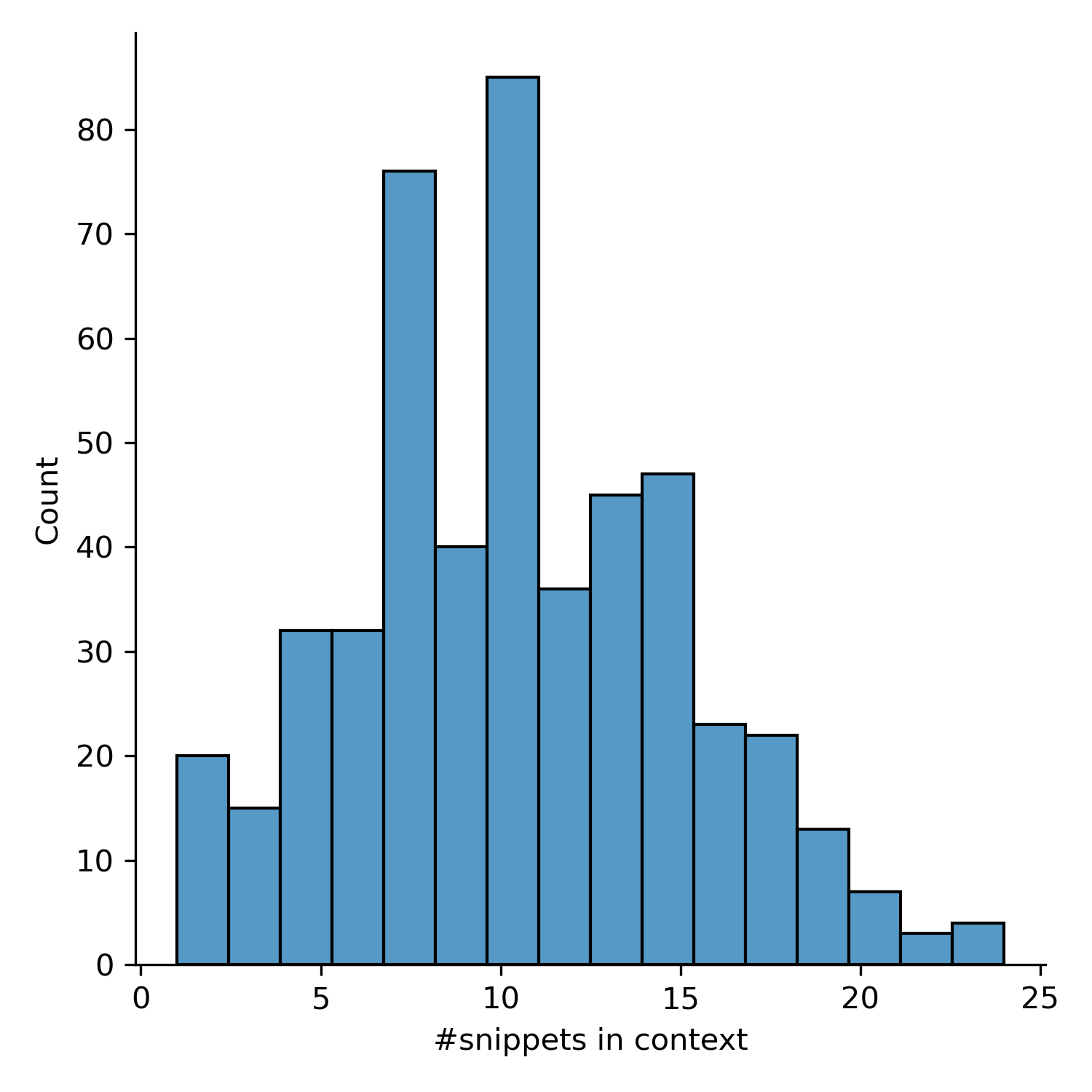}
    \end{subfigure}
      ~ 
    \begin{subfigure}[t]{0.33\textwidth}
    \includegraphics[scale=0.45]{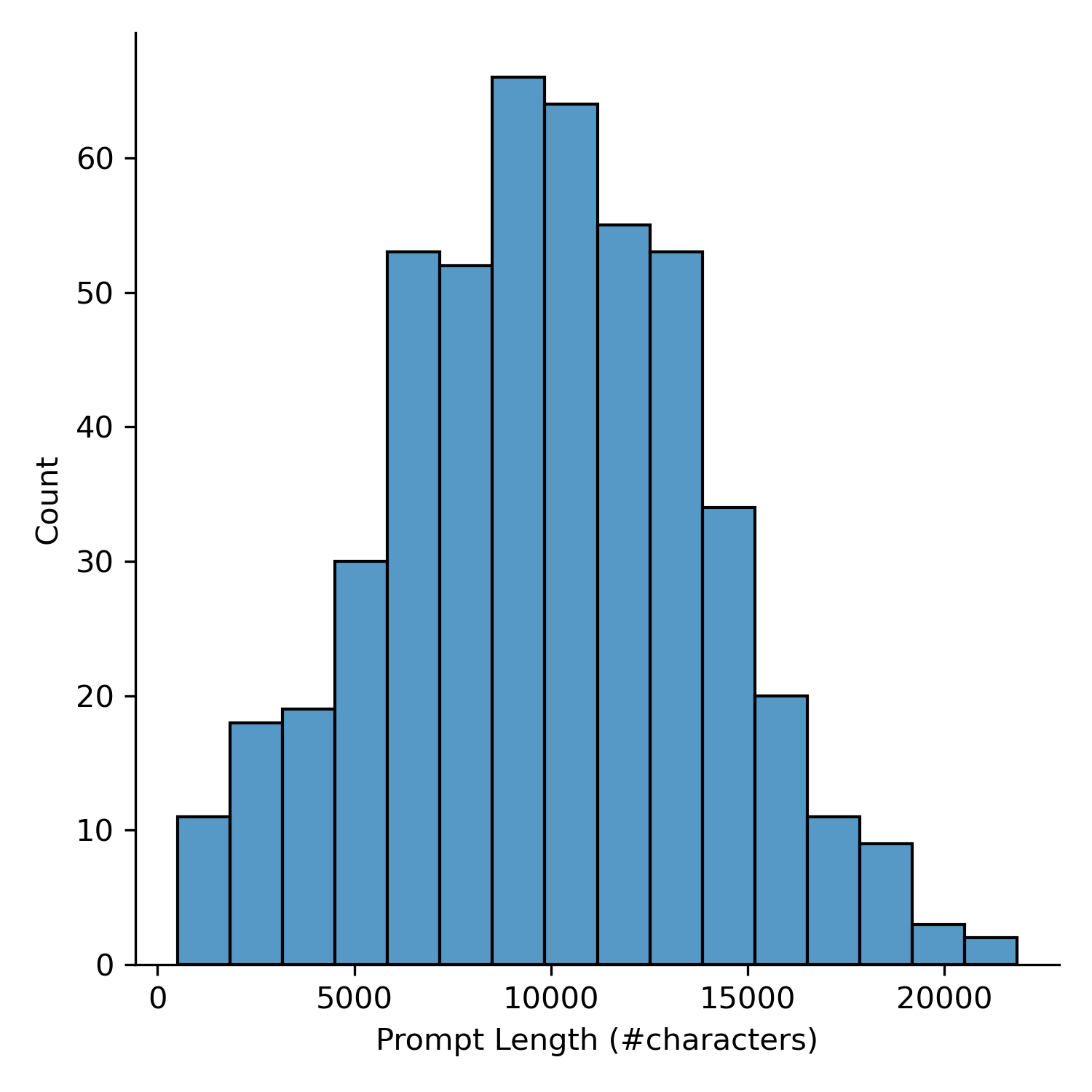}
    \end{subfigure}
    \caption{\label{fig:histogram_numdoc}
 Histogram of number of unique documents (left), number of snippets (middle), and corresponding final prompt length (right) per input context for response generation}
\end{centering}
\end{figure*}

In order to generate the response to the LiveRAG question, the snippets identified during the research phase are used as source material. In particular, the system role is left blank. 
The {\tt{user}} role is the prompt in Figure~\ref{fig:write_prompt}. This prompt has three attributes:
\begin{itemize}
\item  {\tt{context}} - a concatenation of the text in each of the snippets that was accepted after the final filtering stage
\item {\tt{question}} - the original LiveRAG question.
\item {\tt{total\_words}} -  200
\end{itemize}
A local version of Falcon3-10B-Instruct was used to generate the final response as specified
in the rules for LiveRAG. The order of snippets in the prompt is not a rank order. Instead snippet ordering is determined by the rank order of the passage to the individual query that retrieved it as well as the lexical ordering of the chunk in the passage. More precisely, generated query $Q_1$ retrieves passages $A$, $B$ and $C$ in rank order. Generated query $Q_2$ retrieves passages $D$, $E$, and $F$ in rank order. The original question $Q_o$ retrieves passages $G$, $H$, and $I$. Passages are ordered $A$ to $I$. While passages are of length 450 tokens, they have a variable number of characters since not all tokens are of the same length. Each of the passages are divided into $\left\lceil \text{length of} (Passage) / 1000 \right\rceil$ snippets. Some of the snippets are filtered because they are not sufficiently similar to the original query.

In addition to the prompt for answer generation and the generated answer, LiveRAG asked for a list of documents. The documents were ordered based on their retrieval for the individual queries. The three queries were concatenated together. While nine passages were retrieved, there may have been fewer documents if multiple passages from the same document were retrieved by one or more queries.
Only the first time the document occurs is included in the final list.

\section{Experiments before Live Challenge Day}

The LiveRAG system was adapted from the system developed for the NeuCLIR Pilot Report Generation task~\cite{neuclir-coe24}. Prompts for that system were optimized for Claude and GPT-4. Prompts for LiveRAG were adapted to Qwen and Falcon for the Query Generation Prompt and the Response Generation Prompt respectively. 
Development data from DataMorgana was used to generate sample questions that could be used to spot check the system, both for the ability to retrieve documents that were responsive to the query as well as to check that answers reflected the information found in the documents.

The main changes made to the system during development were with an eye towards improving efficiency.
The number of additional generated queries was reduced from four to two.
For each query the number of passages was reduced from four to three.

During the dry run, additional human assessment was undertaken. In general the snippets contained information pertaining to the question and in a vast majority of cases, that information was carried forward to the answer. It was noted that on occasion a snippet contained more specific information that could have yielded a more precise answer than was generated. 


The main modification that was made as the challenge approached was to ensure that the retrieval
system could return responses at a sufficient rate.
Since the generation component expects the content of the retrieved passages from the search engine, PLAID-X service needs to host and serve the body text. 
We identified that the primary latency came from the excessive duplication of the collection in memory due to the multi-thread search service. We modified the service to host the content separately to reduce the latency sufficiently for the live challenge. Each query, after the modification, takes around 400 ms to serve with an NVIDIA V100 GPU. 

To ensure that the system would respond to all 500 questions in two hours, the question file was divided in half and two processes were used. We used a single retrieval service that batched queries for efficiency and a single version of each of the LLMs for generation. We were able to complete all generation in just over one hour.

\section{Automated Results}

The HLTCOE system placed fifth among the participating teams in terms of correctness with a score of 1.070111.
The faithfulness score of 0.340711 was 14th among the 25 teams participating. 

Appendix~\ref{sec:a1} includes Table \ref{tab:example_output} that illustrates an example output from our system in detail. Note that 2 queries were generated based on an initial search, after which 7 documents were retrieved in total. The passages are then chunked into smaller-sized snippets and filtered according to embedding similarity with the queries. From the example, we observe that many snippets are potentially relevant, even if nothing directly answers the question, e.g. (1), (6), (7); there are also a few clearly irrelevant snippets, e.g. (16). 
Note that multiple snippets from the same passage may be included in the context. In this case, we have 16 snippets from the 7 documents. In our system, we observe the final response LLM generally takes a large set of snippets as input. Figure \ref{fig:histogram_numdoc} shows the variation in numbers of documents and snippets as well as the final prompt length provided in each prompt over the 500 query challenge set.

We also performed some analysis after the live challenge. Table \ref{tab:compare_falcon_llama} in Appendix~\ref{sec:a1} compares two different LLMs for response generation. The Falcon-10B Instruct model was used for the challenge, while a larger Llama-70B-Instruct model is compared. Anecdotally, we do not observe a significant difference in the quality of the responses; this suggests that improving the document retrieval and context construction steps may play a more important role than the size and family of model to affect the final outcome of the generation.

\section{Future Work}

In the future we are interested in investigating how the order of content impacts the generation and how aggressive the filtering should be. While the order in which snippets were represented at generation was not strictly based on relevance, it was not clear how best to structure the information in the generation prompt to ensure that the most relevant information would be included in the output. This is worth further investigation. In these experiments, filtering was assessed for each passage independently; however, a model that can reason over more information at once would be able to prioritize information and remove redundant information. This may of interesting implcation for RAG tasks. 

\section{Summary}

In conclusion, the HLTCOE team developed an effective RAG system that incorporated state-of-the-art components. Our search engine was able to identify content that was useful for addressing the LiveRAG questions. Our generation process was effective at proposing additional queries and assembling the information into a meaningful answer. 

\bibliographystyle{ACM-Reference-Format}
\bibliography{sample-base}

\appendix

\section{Examples}
\label{sec:a1}
Table~\ref{tab:example_output} contains example output. Table~\ref{tab:compare_falcon_llama}
compares Falcom to Llama generation.

\begin{table*}
\caption{\label{tab:example_output}
Example system output. Generated queries are the result from the LLM call shown in Figure \ref{fig:research_prompt}. Filtered passages form input context for the LLM call shown in Figure \ref{fig:write_prompt}. }
\begin{small}
    \centering
    \begin{tabular}{|p{0.11\linewidth} | p{0.83\linewidth}|}\hline
        Input Question & I live close to a park with many geese. Can I get bird flu from their droppings?  \\\hline
        Generated Queries & (1) can i get bird flu from geese droppings in 2025; (2) risks of bird flu transmission from geese to humans \\\hline
        Retrieved Docs & 7 documents in total from 3 searches (input question plus 2 generated queries) \\\hline
        Filtered Snippets \hspace{5mm}(docid:snippet) & 
        (1) e7703b83-95fb-497a-859a-c50d8e37e329: "unaffected themselves, the host birds can spread the infection to susceptible species, especially domesticated chickens, turkeys and geese"; 
        (2) ba608cc9-5d39-420c-98ac-b5a3bed83a60: "Droppings dry, then are pulverized in the air or dissolved in the water and from there get to other birds, wild or domestic.";
        (3) e7703b83-95fb-497a-859a-c50d8e37e329: "The outbreak of any epidemic or the widespread of any disease leads to an inevitable slump in the market.";
        (4) ba608cc9-5d39-420c-98ac-b5a3bed83a60: "Sometimes, poultry also caught it. But people have never got sick or died of it.";
        (5) 93b69f5d-b805-4be0-bb73-06be6f33ece4: "Q) Can pets get bird flu? A) Yes.";
        (6) 795b59f7-6303-40dd-ba97-f3872770569c: "Although this virus can spread quickly amongst birds the risk to the public is very low and well cooked eggs and poultry remain safe to eat.";
        (7) 61120bc5-bc10-4c6d-825a-629a0fc13d22: "There are many types of bird flu, and most are harmless to humans, however occasionally, in rare cases where bird flu does infect humans.";
        (8) 93b69f5d-b805-4be0-bb73-06be6f33ece4: "Right now there is a type of bird flu called Highly Pathogenic Avian Influenza (HPAI) type H5N1 (also called H5N1 bird flu) that has made many birds and a few people sick elsewhere in the world.";
        (9) 795b59f7-6303-40dd-ba97-f3872770569c: "a bird keeper in Sulby found 11 of his geese dead near a pond on his property.";
        (10) 61120bc5-bc10-4c6d-825a-629a0fc13d22: "The incubation period for bird flu is 2-8 days which is longer than seasonal flu";
        (11) bd4fd5de-4e73-4875-8f80-6e4c8fd62606: "STP Pest Control Birds are a nuisance creating problems with noise, bird mess, getting in roof spaces and in some cases aggressive behaviour. "
        (12) 93b69f5d-b805-4be0-bb73-06be6f33ece4: "There is no evidence that properly cooked poultry or eggs can be a source of infection from bird flu."
        (13) 4f0151b0-6571-4fa2-8a2f-fd9756f0e3b0: "a hawk enthusiast may be willing to let his birds have at it, so long as the pigeons are not located in a high traffic area that could endanger the hawks."
        (14) bd4fd5de-4e73-4875-8f80-6e4c8fd62606: "Ornithosis Ornithosis is a 'flu-like disease most commonly found in pigeon droppings. Fatalities are rare, but occur as with any other influenza virus.";
        (15) 93b69f5d-b805-4be0-bb73-06be6f33ece4: "Teach children to always wash their hands after playing outside.";
        (16) 4f0151b0-6571-4fa2-8a2f-fd9756f0e3b0: "news of the victory over Napoleon at Waterloo was carried by military homing pigeons" \\
\hline
Answer & Based on the provided sources, there is currently no evidence that properly cooked poultry or eggs can be a source of infection from bird flu. Regarding contact with geese droppings, the sources do not explicitly mention the risk of contracting bird flu from goose droppings in parks or other public places. However, it's generally advised to avoid direct contact with bird droppings due to the potential presence of various bacteria and parasites. To minimize risk, it's recommended to wash hands after being in areas with bird droppings and avoid touching them with bare hands. \\\hline
    \end{tabular}
\end{small}

\end{table*}

\begin{table*}[b!]
\caption{\label{tab:compare_falcon_llama} Comparison of Answers  by Falcon3-10B Instruct (F) vs Llama3.1-70B Instruct (L) when given the same input context}
\begin{small}
    \centering
    \begin{tabular}{|p{0.1in} | p{6.5in}|}\hline

    Q & What's the difference between perspectivism and the despair of defiance regarding human autonomy? \\\hline
    F & Perspectivism posits that our understanding of reality is always mediated by individual perspectives, whereas the despair of defiance, as described by thinkers like Nietzsche, suggests a rejection of traditional moral and metaphysical constraints, leading to a state where one acknowledges the lack of absolute truth or meaning, yet defiantly embraces life's chaos and multiplicity. \\\hline
    L& Perspectivism acknowledges the multiplicity of perspectives and the fragmented nature of human experience, whereas the despair of defiance regarding human autonomy is a more extreme stance that rejects the idea of human autonomy altogether, often in favor of a nihilistic or deterministic worldview. Perspectivism seeks to understand and navigate the complexities of human experience, whereas the despair of defiance tends to abandon hope for human agency and autonomy. \\\hline\hline

    Q &  Which tattoo needle arrangements are most suitable for skin preservation? \\\hline
    F & Based on the provided information, curved magnum (RM) or curved magnum (CM/M1C) tattoo needle arrangements are most suitable for skin preservation. These needles have a curvature that allows for easier movement on the skin, reducing the risk of the needle edges digging in, which is particularly beneficial for delicate areas with curvature, such as the inner elbow. \\\hline
    L& Round curve magnums and curved magnums (RM or CM) are most suitable for skin preservation as they are designed to deflect the skin when moving in and out, reducing the risk of edges digging into the skin and causing damage. These needle arrangements are particularly useful for delicate areas with curvature, such as the inner elbow. \\\hline\hline

    Q & What realms wil be the playble nations in the Europa Universalis board game?  \\\hline
    F & In the Europa Universalis board game, the playable realms will initially focus on European nations, as the game's title and scope suggest a concentration on European history and politics. However, the game may expand to include other regions through future expansions, allowing players to explore and control a broader range of historical territories. \\\hline
    L& The playable nations in the Europa Universalis board game will initially be the European nations, with the main map covering Europe and the Mediterranean region. Future expansions may include other parts of the world, such as North and South America, Africa, Asia, and Australia, but the initial focus will be on European nations. \\\hline
    
    \end{tabular}
    \end{small}
    
\end{table*}


\end{document}